\makeatletter
\declare@file@substitution{revtex4-1.cls}{revtex4-2.cls}
\makeatother
\documentclass[twocolumn, times, tighten,twocolappendix]{aastex63}
\usepackage{amsmath}
\usepackage{amssymb,txfonts}
\usepackage[T1]{fontenc}
\usepackage{graphicx,multirow,hyperref,url,color,xspace}

\newbox\grsign \setbox\grsign=\hbox{$>$} \newdimen\grdimen \grdimen=\ht\grsign
\newbox\simpropbox
\setbox\simpropbox=\hbox{\raise.5ex\hbox{$\propto$}\llap
 {\lower.5ex\hbox{$\sim$}}}\ht2=\grdimen\dp2=0pt

\defcitealias{BZ77}{BZ}
\newcommand{\bz}{{\citetalias{BZ77}}\xspace}

\begin{document}

\title{Formation and evolution of transient jets and their cavities in black-hole X-ray binaries}
\shorttitle{Transient jets}

\author[0000-0002-0333-2452]{Marek Sikora}
\affiliation{Nicolaus Copernicus Astronomical Center, Polish Academy of Sciences, Bartycka 18, PL-00-716 Warszawa, Poland; \href{mailto:sikora@camk.edu.pl}{sikora@camk.edu.pl}}

\author[0000-0002-0333-2452]{Andrzej A. Zdziarski}
\affiliation{Nicolaus Copernicus Astronomical Center, Polish Academy of Sciences, Bartycka 18, PL-00-716 Warszawa, Poland; \href{mailto:sikora@camk.edu.pl}{sikora@camk.edu.pl}}

\shortauthors{Sikora \& Zdziarski}

\begin{abstract}
We propose a model explaining the origin of transient/episodic jets in black-hole X-ray binaries, in which they are caused by transitions from a collimated, strongly magnetized, jet to a wide, un-collimated, outflow. The change occurs when the accretion flow leaves the magnetically-choked state due to an increase of the accretion rate for a weakly varying magnetic flux. The formed powerful jet then detaches from its base, and propagates as a discrete ejection. The uncollimated outflow then produces a relativistic plasma that fills surrounding of the black hole, contributing to the formation of a low-density cavity. While the pressure in the cavity is in equilibrium with the surrounding interstellar medium (ISM), its inertia is orders of magnitude lower than that of the ISM. This implies that the plasma cannot efficiently decelerate the ejecta, explaining most of the observations. The modest deceleration within the cavities observed in some cases can be then due to the presence of clouds and/or filaments, forming a wide transition zone between the cavity and the ISM. 
\end{abstract}

\section{Introduction}
\label{intro}

The concept of launching of powerful jets by rotating black holes (BHs) involving Magnetically Arrested Disks (MAD; \citealt{BK74, Narayan03}) provides new tools to explain a large diversity of accretion flow properties and their time evolution, as well as it explains high jet production efficiencies. It connects the idea of generating initially Poynting-flux dominated jets by rotating black holes (\citealt{BZ77}, hereafter \bz) using the magnetic field of the MAD (\citealt{Igumenshchev08, Tchekhovskoy11,McKinney12} and references therein). That magnetic field is so strong that its pressure counterbalances the ram pressure of the accretion flow. In contrast to standard accretion disc theories, where the structure of accretion flows down to the innermost stable orbits is not affected by magnetic fields and their role is limited to provide viscosity via the magneto-rotational instability, accretion onto a BH in the MAD proceeds via magnetic Rayleigh-Taylor instability across the unipolar flux of poloidal magnetic field threading the BH and innermost parts of the accretion flow (see \citealt{Davis20} for a review).

Independently whether the system reaches the MAD state or not, rotating BHs threaded by poloidal magnetic fields produce Poynting-flux dominated outflows (\bz). Advantages of MAD models over those with weaker fields, named SANE (standard-and-normal-evolution; \citealt{Narayan12b}), are not limited to the possibility of producing most powerful jets observed in AGNs and XRBs, but also by getting them well collimated. Since relativistic \bz outflows cannot self-collimate \citep{Begelman94}, external collimation is needed \citep{Bogovalov99, Globus16}. That collimation can be provided by nonrelativistic (or at most mildly relativistic) MHD outflows from innermost portions of the accretion flow, see, e.g., fig.\ 4 of \citet{McKinney12}. Such outflows are efficient only when the total magnetic flux exceeds the limiting value of the MAD (i.e., the magnetic flux that can be confined on the BH by the ram pressure of the  accretion flow). In that case, the excess poloidal magnetic flux threading the inner accretion flow facilitates the outflows. In the case of SANE models, magnetic disk outflows are either absent or very weak. The \bz BH outflows could be then confined by radiative or thermal accretion disc winds, but those are wide, and not expected to be good collimators on larger scales.

We focus in this paper on production and evolution of transient/episodic jets in BH X-ray binaries (XRBs; e.g., \citealt{FBG04}) and on the nature of cavities surrounding them \citep{Heinz02}. Two main types of jets are observed in BH XRBs, compact and transient. The former are relatively steady jets associated with the hard spectral state (see, e.g., \citealt{ZG04, DGK07} for reviews of spectral states) of BH XRBs. Their sizes observed in radio are limited to $\sim\! 10^{15}$\,cm (e.g., \citealt{Stirling01}) and are often unresolved by radio observations, and thus also referred to as the core jets. The transient jets are associated with transitions from the hard intermediate state (defined as the softest part of the hard state, with the X-ray photon index $>$2) to the soft one, and are observed as discrete moving ejecta \citep{MR94, FBG04}. Often both the approaching and the receding components are seen (e.g., \citealt{Fender99, Bright20}), and they are sometimes detected up to a parsec scale (e.g., \citealt{Carotenuto21}).

This Letter is organized as follows. Our proposed scenario of the production of episodic jets is presented in Section \ref{production}. Their propagation through the cavities and the cavity formation and evolution are investigated in Section \ref{evolution}. Then in Section \ref{discussion}, we discuss our results and compare our results for XRB cavities with those on galactic scales, and in Section \ref{conclusions}, we present our main conclusions.

\section{Production of transient jets}
\label{production}

As we described in Section \ref{intro}, jets produced in the MAD state are effectively collimated, while no such collimation is expected when the magnetic flux drops below the MAD limit. Since the maximum magnetic flux that can thread the BH scales with $\sqrt{\dot M_{\rm acc}}$ (where $\dot M_{\rm acc}$ is the accretion rate), we may expect transitions between strongly collimated (jetted) states and weakly collimated (windy) \bz-outflow states provided variations of $\dot M_{\rm acc}$ lead to changes between $\Phi_{\rm tot}>\Phi_{\rm BH,max}$ (MAD state) and $\Phi_{\rm tot}<\Phi_{\rm BH,max}$ (SANE), where $\Phi_{\rm tot}$ is the net poloidal magnetic flux accumulated on both one hemisphere of the BH and the accretion flow. The maximum magnetic flux that can be confined on the BH by the ram pressure of the accreting plasma is given by \citep{Davis20}
\begin{equation}
\Phi_{\rm BH,max} = \phi (\dot M_{\rm acc} c r_{\rm g}^2)^{1/2}, \quad \phi \approx 70 (1-0.38 a r_{\rm g}/r_{\rm h}) h_{0.3}^{1/2},
\label{phi}
\end{equation}
where $\phi$ is a dimensionless magnetic flux, $h\equiv r\times 0.3 h_{0.3}$ is the half-thickness of the disc at radius $r$, and $a$ is the dimensionless BH spin. The above formula for $\phi$ approximates well results of numerical simulations (\citealt{Davis20} and references therein). 

\begin{figure}
\centerline{ \includegraphics[width=7.5cm]{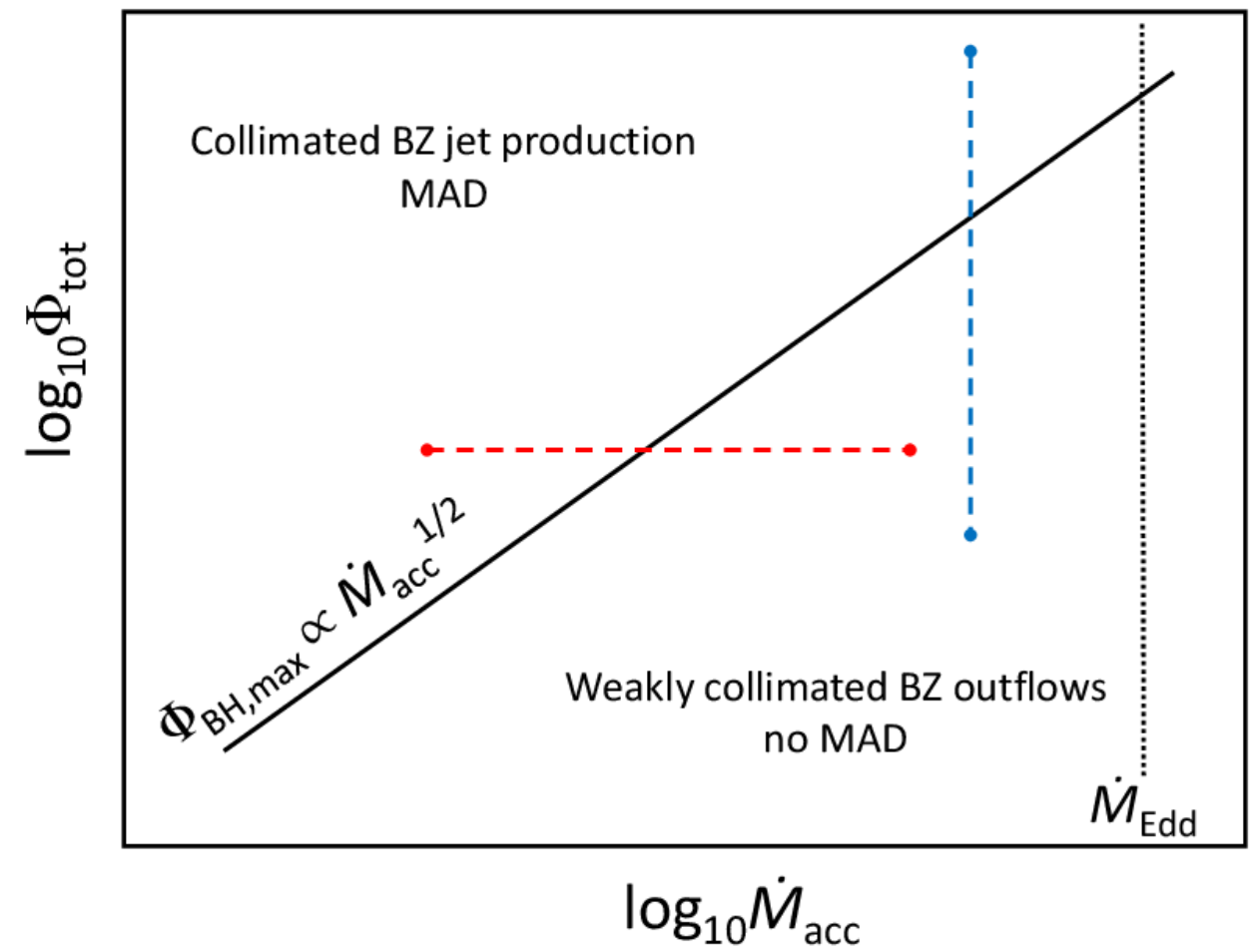}} 
\caption{A schematic illustration of the mechanism driving production of \bz outflows as either jets or winds, based on that in \citet{Rusinek17}. Since the condition for an appearance of MAD is given by $\Phi_{\rm tot}>\Phi_{\rm BH,max}$, where $\Phi_{\rm BH,max}$ is given by equation (\ref{phi}), changes of $\dot M_{\rm acc}$ (and/or the disc scale height, $h/r$) lead to changes between states with jets and with winds, as illustrated by the red dashed line. Furthermore, such changes may occur due to variations of $\Phi_{\rm tot}$, see the blue dashed line. 
} \label{transitions}
\end{figure}

Such transitions are schematically illustrated in Figure \ref{transitions}. Here, XRB states producing jets (both those compact and episodic) are located above the $\Phi_{\rm BH,max}$ line, while those producing wide outflows (in particular those in the soft spectral state) are located below the $\Phi_{\rm BH,max}$ line. We can infer from this diagram and Equation (\ref{phi}) that transitions between jetted and un-jetted states can be driven by changes of the accretion rate, of the total magnetic flux, and of the geometrical thickness of the accretion flow. Our interest here is on production of episodic jets/ejecta observed during the transition from the hard-intermediate state to the soft one. A jet can become episodic when the (total) magnetic flux threading inner parts of the accretion flow is squeezed onto the BH due to an increase of $\dot M_{\rm acc}$. Initially the source is located above the $\Phi_{\rm BH,max}$ line in Figure \ref{transitions}, i.e., in the hard intermediate state. Then an increase of $\dot M_{\rm acc}$ causes it to move to the right, as illustrated by the red dashed line. In this case, the jet power ($\propto \dot M_{\rm acc}$) increases first, but then it crosses the MAD boundary, at which point the \bz outflow is no more collimated and changes into a wide conical Poynting-flux dominated wind. Hence the jet formed up to that point detaches from the BH and continues to move as a discrete ejection. That jet is initially narrow, but it laterally expands during its propagation, and aquires a blobby shape only at large distances. We also note that the total magnetic flux does not need to be constant during a transition. We only need $\Phi_{\rm tot} \propto \dot M^k$ with $k<1/2$ in order to get a transition to the SANE state. Then, such variability of the magnetic flux can explain jet variability observed prior to the ejection (e.g., \citealt{Carotenuto21}). Note that $k$ can be even negative in the case of changing the polarity of poloidal magnetic fields advected to the center (see, e.g., \citealt{Sadowski16}). We note that similar transitions between MAD and SANE states were suggested earlier to explain transitions between radio-loud and radio-quiet quasars \citep{Sikora13, Rusinek17}.

This explains both the powerful episodic ejections and the following absence of core radio emission seen in most transient XRBs in soft state \citep{FBG04, Fender09, Miller-Jones12, Corbel13, Kalemci13}. Alternatively, a change of the accumulated magnetic flux will also cause the outflow to move between the collimated jet and wide outflow states, as illustrated by the blue dashed line. Since the accretion rate during the transition from the hard state to hard intermediate state appears to be associated with an increase of $\dot M_{\rm acc}$, the former case is more likely. In that case we will have a powerful jet whose formation suddenly stops. The amount of energy carried by the ejection is equal to $E=  \Delta t_{\rm inj} P_{\rm j}$, where the ejection duration, $\Delta t_{\rm inj}$, is within a few/several days (e.g., \citealt{Carotenuto22}), and the jet power, $P_{\rm j}$, is close to maximal value achievable by the MAD models \citep{Davis20},
\begin{equation}
P_{\rm j,max}\approx 1.3 h_{0.3} a^2 \dot M_{\rm acc} c^2.
\label{pmax}
\end{equation}

That powerful jet can interact with the remnant of the previous compact jet. Such interaction will be analogous to that in the internal shock model (originally proposed for $\gamma$-ray bursts; \citealt{Rees94}), in which shocks are predicted to be formed as a result of interaction between faster and slower shells. This model has also been proposed for microquasars \citep{Kaiser00, Jamil10, Malzac13, Malzac14}. Then, evolution of the ejecta has the two initial steps.

(1) The less powerful, slower-moving, compact jet forms a classical double-shock structure while interacting with the transient jet  (see, e.g., fig.\ 1 in \citealt{Spada01} but replacing shells by jets). Due to having balance of the momentum fluxes at the contact surface, shocked fluids propagate in this phase 
with the constant Lorentz factor, $\Gamma_{\rm sh}$, between $\Gamma_{\rm f}$ of the faster transient jet and $\Gamma_{\rm s}$ of the slower compact jet.

(2) The transient jet is entirely shocked and hence its momentum flux is not anymore balanced. From this moment on, the shocked plasmas start to decelerate, but at most down to $\Gamma_{\rm s}$.

\section{The nature of the cavities}
\label{evolution}

Tracking of motion of transient jets is critical for determination of their
energetics. In many cases, radio and X-ray observations have tracked the motion of the approaching and receding ejecta up to large distances \citep{MR94, Hjellming95, Tingay95, Fender99, Corbel02, Corbel05, Tomsick03, Kaaret03, Yang10, Yang11, Miller-Jones11, Rushton17, Russell19, Bright20, Espinasse20, Carotenuto21, Bahramian23}. The ejecta are seen to propagate with at most weak deceleration, implying that the BH XRBs reside in low-density environments, or cavities \citep{Heinz02}. The matter density within the cavities is lower than that typical for warm interstellar medium (ISM) by at least a few orders of magnitude (e.g., \citealt{Heinz02, Wang03, Hao09, Carotenuto22, Zdziarski23a}). In only two cases, of XTE J1550--564 and XTE J1348--630, the ejection was observed to cross the boundary of the cavity and quickly decelerate outside of it, see \citet{Steiner12} and \citet{Carotenuto21}, respectively. In both cases, the ISM boundary was at $\sim$0.5\,pc. The sizes of other observed cavities remain unknown, and they could, in principle, be much larger.

The origin of the cavities remains uncertain. They cannot be related to the supernova outbursts that created the BHs, given the issues of the motion of the XRBs through the ISM and the long ages in the case of low-mass XRBs, see \citet{Hao09}. Then, disc winds are not likely to produce the cavities \citep{Hao09}. The cavities could have been carved by the previous jet activity of the systems, but then they would produce narrow tunnels, which are unlikely to remain sustained over long time, as argued by \citet{Hao09}. On the other hand, a very promising mechanism appears to be inflation of a quasi-spherical cavity by relativistic plasmas from {\it both\/} collimated jets \citep{Heinz08, Yoon11} and uncollimated \bz outflows powered by rotating BHs. The latter occur after transition from the hard intermediate state to the soft state, due to $\Phi_{\rm tot} < \Phi_{\rm BH,max}$, see Figure \ref{transitions}. The relativistic plasma is then expected to form a quasi-spherical cavity and have the pressure approximately equal to the pressure of the surrounding ISM \citep{Heinz08}.

In that case, the pressure of the relativistic gas deposited inside the cavity, $p_{\rm cav}$, is equal to $n_{\rm ISM} k T_{\rm ISM}$, where $n_{\rm ISM}$ and $T_{\rm ISM}$ are the ISM number density and temperature, respectively. In the case of a warm ionized ISM, with $n_{\rm ISM} \approx 1$\,cm$^{-3}$ and $T_{\rm ISM}\approx 10^4$\,K, the energy needed to be deposited within a cavity of a given volume equals $E_{\rm cav}=(u_{\rm cav}+p_{\rm cav}) V_{\rm cav}$, where $u_{\rm cav}$ is the energy density. In a spherical case,
\begin{equation}
E_{\rm cav} \approx 4 n_{\rm ISM} k T_{\rm ISM} \frac{4\pi}{3} R_{\rm cav}^3\!\approx 8.5\times 10^{43} \left(\frac{R_{\rm cav}}{0.5\,{\rm pc}}\right)^3{\rm erg},
\label{Ecav}
\end{equation}
where $R_{\rm cav}$ is the cavity radius. The time needed to inflate such a cavity is
\begin{equation}
 t \approx 0.27 \frac {(R_{\rm cav}/ 0.5\,{\rm pc})^3} {P_{\rm j}/10^{37}\, {\rm erg\,s}^{-1}} {\rm yr}.
\end{equation}
This implies that the cavities with the detected boundaries could, in principle, be inflated during single outbursts of XRBs.

In the idealized case of steady-state injection of relativistic plasma into the cavity, we expect the formation of a four-zone structure: (i) the innermost part filled up by quasi-spherical relativistic outflow; (ii) the shocked central outflow; (iii) the shocked external medium; (iv) the un-shocked ISM. In this classical double-shock structure, the cavity edges are represented by discontinuity/contact surface, and, provided there is no motion of the central source relative to ISM, one may expect continuous increase of the cavity.  

While we do not know the cavity sizes for most of the observed cases, it is possible that the two cases with the measured sized (XTE J1550--564 and XTE J1348--630) are representative. Then a question arises how to avoid inflation of a cavity up to $\gg 1$\, pc over multiple outbursts during the lifetime of an XRB. This could be avoided if the plasmas within cavities cool fast enough during quiescent epochs. However, both synchrotron and inverse Compton mechanisms are extremely inefficient. Assuming $B^2/8\pi=p_{\rm cav}$ gives $B\approx 6\,\mu$G, which is somewhat higher than the typical interstellar magnetic field strength. This gives the cooling time of $\sim\! 10^{12}(10^6/\gamma)$\,yr, where $\gamma$ is the electron Lorentz factor. The diluted central accretion emission at the Eddington limit dominates over the starlight, but it still gives inverse-Compton cooling time scale of $\sim\! 10^{9}(10^6/\gamma)$\,yr.

Hence, the only way to avoid too large growth of cavities is to assume that XRBs are moving in the ISM. That motion can be related to the kicks which binaries receive following supernova explosions. As observations show, velocities of such kicks are $v_{\rm kick}\sim 10^2$\,km/s (e.g., \citealt{Atri19}). Then, cavities will form structures extended along the XRB trajectories \citep{Heinz08, Yoon11} like those along the pulsar trajectories (see, e.g., fig.\ 9 in \citealt{Gaensler04}). The BH spins not perpendicular to such trajectories can be responsible for observed in some XRBs asymmetries between the motion of jets and counterjets, e.g., \citet{Hao09}. Such cavities will keep their parsec scale cross-sectional radius provided they were powered with the average rates (including quiescent epochs) equal to
\begin{equation}
\langle P_{\rm j}\rangle \approx 4 p_{\rm cav} \pi R_{\rm cav}^2 v_{\rm kick}\approx 4\times 10^{32} \left(\frac{R_{\rm cav}}{0.5\,{\rm pc}}\right)^2 \frac{v_{\rm kick}}{10^2\,{\rm km/s}} \frac{\rm erg}{\rm s}.
\label{Pj}
\end{equation}

Next, a question emerges whether interactions with the plasma would not stop the ejection at much smaller distances than those observed. In a cold medium, the deceleration is due to the rest energy of ions swept up by the moving ejection (e.g., \citealt{Huang99}). In a hot medium, the deceleration will be due to the total, rest + internal, energy density of the plasma, see Appendix \ref{deceleration} for details. If the plasma is highly relativistic, the internal energy dominates. Then, we can find the density of cold ions with the same total energy density, $u_{\rm p}$, as the total energy density, $u_{\rm cav}$, of the relativistic plasma, and compare it to estimates of the density of ions in observed cavities, which are of the order of $n_{\rm p,obs}\sim 10^{-3}$\,cm$^{-3}$ (e.g., \citealt{Hao09}). Noting that the relativistic plasma in the cavity is in the pressure balance with the external thermal plasma, we obtain $u_{\rm cav}\approx 3 p_{\rm cav} \approx 3 n_{\rm p, ISM} kT_{\rm ISM}$ (where, for the sake of simplicity, we considered only protons). Hence, the number density of protons for which $u_{\rm cav} = u_{\rm p}$ is
\begin{equation}
n_{\rm p} = \frac {3 n_{\rm p,ISM} k T_{\rm ISM}} {m_{\rm p} c^2} \approx 2.8 \times 10^{-9} n_{\rm p, ISM} \frac{T_{\rm ISM}}{10^4\,{\rm K}},
\label{n_p}
\end {equation}
which we normalized using values characteristic to the warm ISM. The same value is obtained for the hot ISM with $n_{\rm ISM} \approx 10^{-4} {\rm cm^{-3}}$ and $T_{\rm ISM} \sim 10^8$\,K. Thus, the relativistic plasma filling the cavity does not noticeably slows down the ejecta. 

At the end of Section \ref{production}, we have shown that the interaction with the remnants of the previous compact jet can only moderately slow down the ejecta and only close the BH. Then, in order to explain the observed wide deceleration profiles of ejecta, we postulate the presence of an extended transition region between the core of the cavity  and the ISM, similarly to the transition layer found in \citet{Zdziarski23a}, but wider and with the  cold protons being associated with clumps/filaments of thermal plasma being
in the  pressure balance with the relativistic plasma. Also, the \bz outflows can be loaded by some baryons, though their contribution to the drag is probably negligible, see Appendix \ref{deceleration}. In particular, in the case of relativistic jets dominated by cold ions in the comoving frame, the ions will be injected to cavities with relativistic energies, $\gamma \sim \Gamma_{\rm j}$, upon the jet complete deceleration.

\section{Discussion}
\label{discussion}

In all previously studies of the evolution of transient jets, the presence of relativistic plasmas inside cavities was not included, despite that inflation of cavities by relativistic plasma from jets was investigated in literature \citep{Heinz08, Yoon11}. As we have argued, the presence of such plasmas in the pressure balance with the external medium is unavoidable. Furthermore, we have demonstrated that the inertia of the relativistic plasma is by several orders of magnitude too low to affect the dynamics of ejecta within the cavities. That may suggest the existence of extended two-phase transition layers, with thermal clumps and/or filaments in the pressure balance with the surrounding relativistic plasma and filling the space near the edges with gradients of the volume filling factor, cf.\ \citet{Zdziarski23a}.
 
We note that the idea of having empty cavities was also considered for X-ray--mapped hot galactic and galaxy-cluster atmospheres. However, radio detections at very low frequencies indicate that the X-ray cavities are not empty, but are filled by relativistic plasmas \citep{Giacintucci11, Birzan20, Capetti22, Plsek23}. Yet another example of remnants of past jet activity represented by cavities filled up by relativistic plasma is provided by two giant bubbles above and below the Milky Way center, the so-called Fermi bubbles \citep{Su10}. They were also detected in radio \citep{Carretti13} and X-rays \citep{Predehl20}, and argued to represent remnants of the past jet activity in Sgr A$^*$ (e.g., \citealt{Guo12, Yang22}).

\section{Conclusions}
\label{conclusions}

Our main results are as follows.

We have proposed a novel scenario explaining the origin of transient/episodic jets in BH XRBs. It involves transition from a collimated \bz outflow (jet) to a wide, un-collimated, one due to an increase of the accretion rate. That increase results in an increase of the possible magnetic flux threading the BH, which in turn results in a transition from the MAD phase to the SANE one of the outflow. Consequently, the jet production stops, resulting in a discrete ejection, detached from the central BH.

The \bz outflow continues after the jet ejection, but it is no more collimated. This produces most of the relativistic plasma contributing to the formation of cavities surrounding XRBs. The cavities are not vacuous, but filled by that plasma. 

The relativistic plasma filling a cavity is in pressure equilibrium with the surrounding ISM. We show that this implies that the plasma cannot decelerate the ejecta. The modest deceleration within the cavities observed in a number of cases requires the presence of clouds and/or filaments filling the cavity, and forming a wide transition layer between the cavity and the surrounding ISM. 

\section*{Acknowledgements}
We thank the referee for valuable comments. We acknowledge support from the Polish National Science Center under the grants 2019/35/B/ST9/03944 and 2017/27/B/ST9/01940.

\appendix
\section{Deceleration by relativistic plasma}
\label{deceleration}

The drag force acting on an ejection moving in an external medium is equivalent to the momentum flux of the medium measured in the ejection comoving frame (e.g., \citealt{Rosen99}). We find it is equal to 
\begin{equation}
\Pi' = (w_{\rm cav} \beta_{\rm ej}^2 \Gamma_{\rm ej}^2 + p_{\rm cav}) A_{\rm ej},
\label{momentum}
\end{equation}
where $w_{\rm cav}=\rho_{\rm cav} c^2 + u_{\rm cav} + p_{\rm cav}$, $\beta_{\rm ej} = v_{\rm ej}/c$, $\Gamma_{\rm ej}= (1-\beta_{\rm ej}^2)^{-1/2}$, and $A_{\rm ej}$ is the jet cross-sectional area. For a cavity filled with a relativistic plasma, we have $\rho_{\rm cav}c^2 \ll u_{\rm cav} + p_{\rm cav}$ and $w_{\rm rel} \approx 4 p_{\rm cav}$, implying 
\begin{equation}
\Pi'_{\rm rel}/A_{\rm ej} \approx 4 p_{\rm cav} \beta_{\rm ej}^2 \Gamma_{\rm ej}^2 + p_{\rm cav} \approx  4 p_{\rm cav} \Gamma_{\rm ej}^2.
\end{equation}
For a cavity filled with a cold plasma, we have
$w_{\rm cav} \approx \rho_{\rm cav}c^2 = n_{\rm p} m_{\rm p} c^2$, and
\begin{equation}
  \Pi'_{\rm cold}/A_{\rm ej} \approx n_{\rm p} m_{\rm p} c^2 \beta_{\rm ej} \Gamma_{\rm ej}^2
  \approx n_{\rm p} m_{\rm p} c^2 \Gamma_{\rm ej}^2.
\end{equation}
(In this case at $\beta_{\rm ej} \ll 1$, we get the well-known formula
for the ram pressure $p_{\rm ram} \equiv  \Pi'_{\rm cold}/ A_{\rm ej} = \rho_{\rm cav} v_{\rm ej}^2$.)

Consequently, the pressure balance between the relativistic plasma in cavity and the pressure of the thermal plasma in the ISM outside it implies $\Pi'_{\rm rel} \ll \Pi'_{\rm cold}$. Then, as discussed in Section \ref{evolution}, such relativistic plasma cannot significantly decelerate the ejecta within the observed cavities, see Equation (\ref{n_p}), and we need another cavity components for that.

\bibliographystyle{aasjournal}
\bibliography{../allbib} 

\begin{thebibliography}{}
\expandafter\ifx\csname natexlab\endcsname\relax\def\natexlab#1{#1}\fi
\providecommand{\url}[1]{\href{#1}{#1}}
\providecommand{\dodoi}[1]{doi:~\href{http://doi.org/#1}{\nolinkurl{#1}}}
\providecommand{\doeprint}[1]{\href{http://ascl.net/#1}{\nolinkurl{http://ascl.net/#1}}}
\providecommand{\doarXiv}[1]{\href{https://arxiv.org/abs/#1}{\nolinkurl{https://arxiv.org/abs/#1}}}

\bibitem[{{Atri} {et~al.}(2019){Atri}, {Miller-Jones}, {Bahramian}, {Plotkin},
  {Jonker}, {Nelemans}, {Maccarone}, {Sivakoff}, {Deller}, {Chaty}, {Torres},
  {Horiuchi}, {McCallum}, {Natusch}, {Phillips}, {Stevens}, \&
  {Weston}}]{Atri19}
{Atri}, P., {Miller-Jones}, J.~C.~A., {Bahramian}, A., {et~al.} 2019, \mnras,
  489, 3116, \dodoi{10.1093/mnras/stz2335}

\bibitem[{{Bahramian} {et~al.}(2023){Bahramian}, {Tremou}, {Tetarenko},
  {Miller-Jones}, {Fender}, {Corbel}, {Williams}, {Strader}, {Carotenuto},
  {Salinas}, {Kennea}, {Motta}, {Woudt}, {Matthews}, \&
  {Russell}}]{Bahramian23}
{Bahramian}, A., {Tremou}, E., {Tetarenko}, A.~J., {et~al.} 2023, \apjl, 948,
  L7, \dodoi{10.3847/2041-8213/accde1}

\bibitem[{{Begelman} \& {Li}(1994)}]{Begelman94}
{Begelman}, M.~C., \& {Li}, Z.-Y. 1994, \apj, 426, 269, \dodoi{10.1086/174061}

\bibitem[{{B{\^\i}rzan} {et~al.}(2020){B{\^\i}rzan}, {Rafferty}, {Br{\"u}ggen},
  {Botteon}, {Brunetti}, {Cuciti}, {Edge}, {Morganti}, {R{\"o}ttgering}, \&
  {Shimwell}}]{Birzan20}
{B{\^\i}rzan}, L., {Rafferty}, D.~A., {Br{\"u}ggen}, M., {et~al.} 2020, \mnras,
  496, 2613, \dodoi{10.1093/mnras/staa1594}

\bibitem[{{Bisnovatyi-Kogan} \& {Ruzmaikin}(1974)}]{BK74}
{Bisnovatyi-Kogan}, G.~S., \& {Ruzmaikin}, A.~A. 1974, \apss, 28, 45,
  \dodoi{10.1007/BF00642237}

\bibitem[{{Blandford} \& {Znajek}(1977)}]{BZ77}
{Blandford}, R.~D., \& {Znajek}, R.~L. 1977, \mnras, 179, 433,
  \dodoi{10.1093/mnras/179.3.433}

\bibitem[{{Bogovalov} \& {Tsinganos}(1999)}]{Bogovalov99}
{Bogovalov}, S., \& {Tsinganos}, K. 1999, \mnras, 305, 211,
  \dodoi{10.1046/j.1365-8711.1999.02413.x}

\bibitem[{{Bright} {et~al.}(2020){Bright}, {Fender}, {Motta}, {Williams},
  {Moldon}, {Plotkin}, {Miller-Jones}, {Heywood}, {Tremou}, {Beswick},
  {Sivakoff}, {Corbel}, {Buckley}, {Homan}, {Gallo}, {Tetarenko}, {Russell},
  {Green}, {Titterington}, {Woudt}, {Armstrong}, {Groot}, {Horesh}, {van der
  Horst}, {K{\"o}rding}, {McBride}, {Rowlinson}, \& {Wijers}}]{Bright20}
{Bright}, J.~S., {Fender}, R.~P., {Motta}, S.~E., {et~al.} 2020, Nature
  Astronomy, 4, 697, \dodoi{10.1038/s41550-020-1023-5}

\bibitem[{{Capetti} {et~al.}(2022){Capetti}, {Brienza}, {Balmaverde}, {Best},
  {Baldi}, {Drabent}, {G{\"u}rkan}, {Rottgering}, {Tasse}, \&
  {Webster}}]{Capetti22}
{Capetti}, A., {Brienza}, M., {Balmaverde}, B., {et~al.} 2022, \aap, 660, A93,
  \dodoi{10.1051/0004-6361/202142911}

\bibitem[{{Carotenuto} {et~al.}(2022){Carotenuto}, {Tetarenko}, \&
  {Corbel}}]{Carotenuto22}
{Carotenuto}, F., {Tetarenko}, A.~J., \& {Corbel}, S. 2022, \mnras, 511, 4826,
  \dodoi{10.1093/mnras/stac329}

\bibitem[{{Carotenuto} {et~al.}(2021){Carotenuto}, {Corbel}, {Tremou},
  {Russell}, {Tzioumis}, {Fender}, {Woudt}, {Motta}, {Miller-Jones}, {Chauhan},
  {Tetarenko}, {Sivakoff}, {Heywood}, {Horesh}, {van der Horst}, {Koerding}, \&
  {Mooley}}]{Carotenuto21}
{Carotenuto}, F., {Corbel}, S., {Tremou}, E., {et~al.} 2021, \mnras, 504, 444,
  \dodoi{10.1093/mnras/stab864}

\bibitem[{{Carretti} {et~al.}(2013){Carretti}, {Crocker}, {Staveley-Smith},
  {Haverkorn}, {Purcell}, {Gaensler}, {Bernardi}, {Kesteven}, \&
  {Poppi}}]{Carretti13}
{Carretti}, E., {Crocker}, R.~M., {Staveley-Smith}, L., {et~al.} 2013, \nat,
  493, 66, \dodoi{10.1038/nature11734}

\bibitem[{{Corbel} {et~al.}(2013){Corbel}, {Coriat}, {Brocksopp}, {Tzioumis},
  {Fender}, {Tomsick}, {Buxton}, \& {Bailyn}}]{Corbel13}
{Corbel}, S., {Coriat}, M., {Brocksopp}, C., {et~al.} 2013, \mnras, 428, 2500,
  \dodoi{10.1093/mnras/sts215}

\bibitem[{{Corbel} {et~al.}(2002){Corbel}, {Fender}, {Tzioumis}, {Tomsick},
  {Orosz}, {Miller}, {Wijnands}, \& {Kaaret}}]{Corbel02}
{Corbel}, S., {Fender}, R.~P., {Tzioumis}, A.~K., {et~al.} 2002, Science, 298,
  196, \dodoi{10.1126/science.1075857}

\bibitem[{{Corbel} {et~al.}(2005){Corbel}, {Kaaret}, {Fender}, {Tzioumis},
  {Tomsick}, \& {Orosz}}]{Corbel05}
{Corbel}, S., {Kaaret}, P., {Fender}, R.~P., {et~al.} 2005, \apj, 632, 504,
  \dodoi{10.1086/432499}

\bibitem[{{Davis} \& {Tchekhovskoy}(2020)}]{Davis20}
{Davis}, S.~W., \& {Tchekhovskoy}, A. 2020, \araa, 58, 407,
  \dodoi{10.1146/annurev-astro-081817-051905}

\bibitem[{{Done} {et~al.}(2007){Done}, {Gierli{\'n}ski}, \& {Kubota}}]{DGK07}
{Done}, C., {Gierli{\'n}ski}, M., \& {Kubota}, A. 2007, \aapr, 15, 1,
  \dodoi{10.1007/s00159-007-0006-1}

\bibitem[{{Espinasse} {et~al.}(2020){Espinasse}, {Corbel}, {Kaaret}, {Tremou},
  {Migliori}, {Plotkin}, {Bright}, {Tomsick}, {Tzioumis}, {Fender}, {Orosz},
  {Gallo}, {Homan}, {Jonker}, {Miller-Jones}, {Russell}, \&
  {Motta}}]{Espinasse20}
{Espinasse}, M., {Corbel}, S., {Kaaret}, P., {et~al.} 2020, \apjl, 895, L31,
  \dodoi{10.3847/2041-8213/ab88b6}

\bibitem[{{Fender} {et~al.}(2004){Fender}, {Belloni}, \& {Gallo}}]{FBG04}
{Fender}, R.~P., {Belloni}, T.~M., \& {Gallo}, E. 2004, \mnras, 355, 1105,
  \dodoi{10.1111/j.1365-2966.2004.08384.x}

\bibitem[{{Fender} {et~al.}(1999){Fender}, {Garrington}, {McKay}, {Muxlow},
  {Pooley}, {Spencer}, {Stirling}, \& {Waltman}}]{Fender99}
{Fender}, R.~P., {Garrington}, S.~T., {McKay}, D.~J., {et~al.} 1999, \mnras,
  304, 865, \dodoi{10.1046/j.1365-8711.1999.02364.x}

\bibitem[{{Fender} {et~al.}(2009){Fender}, {Homan}, \& {Belloni}}]{Fender09}
{Fender}, R.~P., {Homan}, J., \& {Belloni}, T.~M. 2009, \mnras, 396, 1370,
  \dodoi{10.1111/j.1365-2966.2009.14841.x}

\bibitem[{{Gaensler} {et~al.}(2004){Gaensler}, {van der Swaluw}, {Camilo},
  {Kaspi}, {Baganoff}, {Yusef-Zadeh}, \& {Manchester}}]{Gaensler04}
{Gaensler}, B.~M., {van der Swaluw}, E., {Camilo}, F., {et~al.} 2004, \apj,
  616, 383, \dodoi{10.1086/424906}

\bibitem[{{Giacintucci} {et~al.}(2011){Giacintucci}, {O'Sullivan}, {Vrtilek},
  {David}, {Raychaudhury}, {Venturi}, {Athreya}, {Clarke}, {Murgia},
  {Mazzotta}, {Gitti}, {Ponman}, {Ishwara-Chandra}, {Jones}, \&
  {Forman}}]{Giacintucci11}
{Giacintucci}, S., {O'Sullivan}, E., {Vrtilek}, J., {et~al.} 2011, \apj, 732,
  95, \dodoi{10.1088/0004-637X/732/2/95}

\bibitem[{{Globus} \& {Levinson}(2016)}]{Globus16}
{Globus}, N., \& {Levinson}, A. 2016, \mnras, 461, 2605,
  \dodoi{10.1093/mnras/stw1474}

\bibitem[{{Guo} \& {Mathews}(2012)}]{Guo12}
{Guo}, F., \& {Mathews}, W.~G. 2012, \apj, 756, 181,
  \dodoi{10.1088/0004-637X/756/2/181}

\bibitem[{{Hao} \& {Zhang}(2009)}]{Hao09}
{Hao}, J.~F., \& {Zhang}, S.~N. 2009, \apj, 702, 1648,
  \dodoi{10.1088/0004-637X/702/2/1648}

\bibitem[{{Heinz}(2002)}]{Heinz02}
{Heinz}, S. 2002, \aap, 388, L40, \dodoi{10.1051/0004-6361:20020402}

\bibitem[{{Heinz} {et~al.}(2008){Heinz}, {Grimm}, {Sunyaev}, \&
  {Fender}}]{Heinz08}
{Heinz}, S., {Grimm}, H.~J., {Sunyaev}, R.~A., \& {Fender}, R.~P. 2008, \apj,
  686, 1145, \dodoi{10.1086/591435}

\bibitem[{{Hjellming} \& {Rupen}(1995)}]{Hjellming95}
{Hjellming}, R.~M., \& {Rupen}, M.~P. 1995, \nat, 375, 464,
  \dodoi{10.1038/375464a0}

\bibitem[{{Huang} {et~al.}(1999){Huang}, {Dai}, \& {Lu}}]{Huang99}
{Huang}, Y.~F., {Dai}, Z.~G., \& {Lu}, T. 1999, \mnras, 309, 513,
  \dodoi{10.1046/j.1365-8711.1999.02887.x}

\bibitem[{{Igumenshchev}(2008)}]{Igumenshchev08}
{Igumenshchev}, I.~V. 2008, \apj, 677, 317, \dodoi{10.1086/529025}

\bibitem[{{Jamil} {et~al.}(2010){Jamil}, {Fender}, \& {Kaiser}}]{Jamil10}
{Jamil}, O., {Fender}, R.~P., \& {Kaiser}, C.~R. 2010, \mnras, 401, 394,
  \dodoi{10.1111/j.1365-2966.2009.15652.x}

\bibitem[{{Kaaret} {et~al.}(2003){Kaaret}, {Corbel}, {Tomsick}, {Fender},
  {Miller}, {Orosz}, {Tzioumis}, \& {Wijnands}}]{Kaaret03}
{Kaaret}, P., {Corbel}, S., {Tomsick}, J.~A., {et~al.} 2003, \apj, 582, 945,
  \dodoi{10.1086/344540}

\bibitem[{{Kaiser} {et~al.}(2000){Kaiser}, {Sunyaev}, \& {Spruit}}]{Kaiser00}
{Kaiser}, C.~R., {Sunyaev}, R., \& {Spruit}, H.~C. 2000, \aap, 356, 975,
  \dodoi{10.48550/arXiv.astro-ph/0001501}

\bibitem[{{Kalemci} {et~al.}(2013){Kalemci}, {Din{\c{c}}er}, {Tomsick},
  {Buxton}, {Bailyn}, \& {Chun}}]{Kalemci13}
{Kalemci}, E., {Din{\c{c}}er}, T., {Tomsick}, J.~A., {et~al.} 2013, \apj, 779,
  95, \dodoi{10.1088/0004-637X/779/2/95}

\bibitem[{{Malzac}(2013)}]{Malzac13}
{Malzac}, J. 2013, \mnras, 429, L20, \dodoi{10.1093/mnrasl/sls017}

\bibitem[{{Malzac}(2014)}]{Malzac14}
---. 2014, \mnras, 443, 299, \dodoi{10.1093/mnras/stu1144}

\bibitem[{{McKinney} {et~al.}(2012){McKinney}, {Tchekhovskoy}, \&
  {Blandford}}]{McKinney12}
{McKinney}, J.~C., {Tchekhovskoy}, A., \& {Blandford}, R.~D. 2012, \mnras, 423,
  3083, \dodoi{10.1111/j.1365-2966.2012.21074.x}

\bibitem[{{Miller-Jones} {et~al.}(2011){Miller-Jones}, {Jonker}, {Ratti},
  {Torres}, {Brocksopp}, {Yang}, \& {Morrell}}]{Miller-Jones11}
{Miller-Jones}, J.~C.~A., {Jonker}, P.~G., {Ratti}, E.~M., {et~al.} 2011,
  \mnras, 415, 306, \dodoi{10.1111/j.1365-2966.2011.18704.x}

\bibitem[{{Miller-Jones} {et~al.}(2012){Miller-Jones}, {Sivakoff},
  {Altamirano}, {Coriat}, {Corbel}, {Dhawan}, {Krimm}, {Remillard}, {Rupen},
  {Russell}, {Fender}, {Heinz}, {K{\"o}rding}, {Maitra}, {Markoff}, {Migliari},
  {Sarazin}, \& {Tudose}}]{Miller-Jones12}
{Miller-Jones}, J.~C.~A., {Sivakoff}, G.~R., {Altamirano}, D., {et~al.} 2012,
  \mnras, 421, 468, \dodoi{10.1111/j.1365-2966.2011.20326.x}

\bibitem[{{Mirabel} \& {Rodr{\'{\i}}guez}(1994)}]{MR94}
{Mirabel}, I.~F., \& {Rodr{\'{\i}}guez}, L.~F. 1994, \nat, 371, 46,
  \dodoi{10.1038/371046a0}

\bibitem[{{Narayan} {et~al.}(2003){Narayan}, {Igumenshchev}, \&
  {Abramowicz}}]{Narayan03}
{Narayan}, R., {Igumenshchev}, I.~V., \& {Abramowicz}, M.~A. 2003, \pasj, 55,
  L69, \dodoi{10.1093/pasj/55.6.L69}

\bibitem[{{Narayan} {et~al.}(2012){Narayan}, {S{\k{a}}dowski}, {Penna}, \&
  {Kulkarni}}]{Narayan12b}
{Narayan}, R., {S{\k{a}}dowski}, A., {Penna}, R.~F., \& {Kulkarni}, A.~K. 2012,
  \mnras, 426, 3241, \dodoi{10.1111/j.1365-2966.2012.22002.x}

\bibitem[{{Pl{\v{s}}ek} {et~al.}(2023){Pl{\v{s}}ek}, {Werner}, {Topinka}, \&
  {Simionescu}}]{Plsek23}
{Pl{\v{s}}ek}, T., {Werner}, N., {Topinka}, M., \& {Simionescu}, A. 2023, arXiv
  e-prints, arXiv:2304.05457, \dodoi{10.48550/arXiv.2304.05457}

\bibitem[{{Predehl} {et~al.}(2020){Predehl}, {Sunyaev}, {Becker}, {Brunner},
  {Burenin}, {Bykov}, {Cherepashchuk}, {Chugai}, {Churazov}, {Doroshenko},
  {Eismont}, {Freyberg}, {Gilfanov}, {Haberl}, {Khabibullin}, {Krivonos},
  {Maitra}, {Medvedev}, {Merloni}, {Nandra}, {Nazarov}, {Pavlinsky}, {Ponti},
  {Sanders}, {Sasaki}, {Sazonov}, {Strong}, \& {Wilms}}]{Predehl20}
{Predehl}, P., {Sunyaev}, R.~A., {Becker}, W., {et~al.} 2020, \nat, 588, 227,
  \dodoi{10.1038/s41586-020-2979-0}

\bibitem[{{Rees} \& {Meszaros}(1994)}]{Rees94}
{Rees}, M.~J., \& {Meszaros}, P. 1994, \apjl, 430, L93, \dodoi{10.1086/187446}

\bibitem[{{Rosen} {et~al.}(1999){Rosen}, {Hughes}, {Duncan}, \&
  {Hardee}}]{Rosen99}
{Rosen}, A., {Hughes}, P.~A., {Duncan}, G.~C., \& {Hardee}, P.~E. 1999, \apj,
  516, 729, \dodoi{10.1086/307143}

\bibitem[{{Rushton} {et~al.}(2017){Rushton}, {Miller-Jones}, {Curran},
  {Sivakoff}, {Rupen}, {Paragi}, {Spencer}, {Yang}, {Altamirano}, {Belloni},
  {Fender}, {Krimm}, {Maitra}, {Migliari}, {Russell}, {Russell}, {Soria}, \&
  {Tudose}}]{Rushton17}
{Rushton}, A.~P., {Miller-Jones}, J.~C.~A., {Curran}, P.~A., {et~al.} 2017,
  \mnras, 468, 2788, \dodoi{10.1093/mnras/stx526}

\bibitem[{{Rusinek} {et~al.}(2017){Rusinek}, {Sikora}, {Kozie{\l}-Wierzbowska},
  \& {Godfrey}}]{Rusinek17}
{Rusinek}, K., {Sikora}, M., {Kozie{\l}-Wierzbowska}, D., \& {Godfrey}, L.
  2017, \mnras, 466, 2294, \dodoi{10.1093/mnras/stw3330}

\bibitem[{{Russell} {et~al.}(2019){Russell}, {Tetarenko}, {Miller-Jones},
  {Sivakoff}, {Parikh}, {Rapisarda}, {Wijnands}, {Corbel}, {Tremou},
  {Altamirano}, {Baglio}, {Ceccobello}, {Degenaar}, {van den Eijnden},
  {Fender}, {Heywood}, {Krimm}, {Lucchini}, {Markoff}, {Russell}, {Soria}, \&
  {Woudt}}]{Russell19}
{Russell}, T.~D., {Tetarenko}, A.~J., {Miller-Jones}, J.~C.~A., {et~al.} 2019,
  \apj, 883, 198, \dodoi{10.3847/1538-4357/ab3d36}

\bibitem[{{Sikora} \& {Begelman}(2013)}]{Sikora13}
{Sikora}, M., \& {Begelman}, M.~C. 2013, \apjl, 764, L24,
  \dodoi{10.1088/2041-8205/764/2/L24}

\bibitem[{{S{\k{a}}dowski}(2016)}]{Sadowski16}
{S{\k{a}}dowski}, A. 2016, \mnras, 462, 960, \dodoi{10.1093/mnras/stw1852}

\bibitem[{{Spada} {et~al.}(2001){Spada}, {Ghisellini}, {Lazzati}, \&
  {Celotti}}]{Spada01}
{Spada}, M., {Ghisellini}, G., {Lazzati}, D., \& {Celotti}, A. 2001, \mnras,
  325, 1559, \dodoi{10.1046/j.1365-8711.2001.04557.x}

\bibitem[{{Steiner} \& {McClintock}(2012)}]{Steiner12}
{Steiner}, J.~F., \& {McClintock}, J.~E. 2012, \apj, 745, 136,
  \dodoi{10.1088/0004-637X/745/2/136}

\bibitem[{{Stirling} {et~al.}(2001){Stirling}, {Spencer}, {de la Force},
  {Garrett}, {Fender}, \& {Ogley}}]{Stirling01}
{Stirling}, A.~M., {Spencer}, R.~E., {de la Force}, C.~J., {et~al.} 2001,
  \mnras, 327, 1273, \dodoi{10.1046/j.1365-8711.2001.04821.x}

\bibitem[{{Su} {et~al.}(2010){Su}, {Slatyer}, \& {Finkbeiner}}]{Su10}
{Su}, M., {Slatyer}, T.~R., \& {Finkbeiner}, D.~P. 2010, \apj, 724, 1044,
  \dodoi{10.1088/0004-637X/724/2/1044}

\bibitem[{{Tchekhovskoy} {et~al.}(2011){Tchekhovskoy}, {Narayan}, \&
  {McKinney}}]{Tchekhovskoy11}
{Tchekhovskoy}, A., {Narayan}, R., \& {McKinney}, J.~C. 2011, \mnras, 418, L79,
  \dodoi{10.1111/j.1745-3933.2011.01147.x}

\bibitem[{{Tingay} {et~al.}(1995){Tingay}, {Jauncey}, {Preston}, {Reynolds},
  {Meier}, {Murphy}, {Tzioumis}, {McKay}, {Kesteven}, {Lovell},
  {Campbell-Wilson}, {Ellingsen}, {Gough}, {Hunstead}, {Jonos}, {McCulloch},
  {Migenes}, {Quick}, {Sinclair}, \& {Smits}}]{Tingay95}
{Tingay}, S.~J., {Jauncey}, D.~L., {Preston}, R.~A., {et~al.} 1995, \nat, 374,
  141, \dodoi{10.1038/374141a0}

\bibitem[{{Tomsick} {et~al.}(2003){Tomsick}, {Corbel}, {Fender}, {Miller},
  {Orosz}, {Tzioumis}, {Wijnands}, \& {Kaaret}}]{Tomsick03}
{Tomsick}, J.~A., {Corbel}, S., {Fender}, R., {et~al.} 2003, \apj, 582, 933,
  \dodoi{10.1086/344703}

\bibitem[{{Wang} {et~al.}(2003){Wang}, {Dai}, \& {Lu}}]{Wang03}
{Wang}, X.~Y., {Dai}, Z.~G., \& {Lu}, T. 2003, \apj, 592, 347,
  \dodoi{10.1086/375638}

\bibitem[{{Yang} {et~al.}(2022){Yang}, {Ruszkowski}, \& {Zweibel}}]{Yang22}
{Yang}, H. Y.~K., {Ruszkowski}, M., \& {Zweibel}, E.~G. 2022, Nature Astronomy,
  6, 584, \dodoi{10.1038/s41550-022-01618-x}

\bibitem[{{Yang} {et~al.}(2010){Yang}, {Brocksopp}, {Corbel}, {Paragi},
  {Tzioumis}, \& {Fender}}]{Yang10}
{Yang}, J., {Brocksopp}, C., {Corbel}, S., {et~al.} 2010, \mnras, 409, L64,
  \dodoi{10.1111/j.1745-3933.2010.00948.x}

\bibitem[{{Yang} {et~al.}(2011){Yang}, {Paragi}, {Corbel}, {Gurvits},
  {Campbell}, \& {Brocksopp}}]{Yang11}
{Yang}, J., {Paragi}, Z., {Corbel}, S., {et~al.} 2011, \mnras, 418, L25,
  \dodoi{10.1111/j.1745-3933.2011.01136.x}

\bibitem[{{Yoon} {et~al.}(2011){Yoon}, {Morsony}, {Heinz}, {Wiersema},
  {Fender}, {Russell}, \& {Sunyaev}}]{Yoon11}
{Yoon}, D., {Morsony}, B., {Heinz}, S., {et~al.} 2011, \apj, 742, 25,
  \dodoi{10.1088/0004-637X/742/1/25}

\bibitem[{{Zdziarski} \& {Gierli{\'n}ski}(2004)}]{ZG04}
{Zdziarski}, A.~A., \& {Gierli{\'n}ski}, M. 2004, Progr. Theor. Phys. Suppl.,
  155, 99, \dodoi{10.1143/PTPS.155.99}

\bibitem[{{Zdziarski} {et~al.}(2023){Zdziarski}, {Sikora}, {Szanecki}, \&
  {B{\"o}ttcher}}]{Zdziarski23a}
{Zdziarski}, A.~A., {Sikora}, M., {Szanecki}, M., \& {B{\"o}ttcher}, M. 2023,
  \apjl, 947, L32, \dodoi{10.3847/2041-8213/accb5a}

\end{thebibliography}

\end{document}